\documentclass[12pt,usletter]{article}

\usepackage{setspace}
\usepackage{color}

\usepackage{amsmath}
\usepackage[dvips]{graphicx}
\usepackage{afterpage}
\usepackage{placeins}
\newcommand{\vect}[1]{\mathbf{#1}}
\newcommand{\diverg}[1]{\vect{\nabla}\cdot\vect{#1}}
\newcommand{\grad}[1]{\vect{\nabla}{#1}}

\renewcommand{\title}[1]{
  \vspace{25mm}
  \noindent {\Large\textbf{#1}\\}
  \vspace{20mm}
}

\renewcommand{\author}[1]{
  \vspace{3mm}
  \noindent \textbf{#1}\\
  \vspace{3mm}
}

\newcommand{\address}[1]{
  \vspace{3mm}
  \noindent #1\par
  \vspace{3mm}
}
\begin{document}
\begin{titlepage}\begin{center}
\title{Simulating Van der Waals-interactions in water/hydrocarbon-based complex fluids}
\author{I. Pasichnyk} 
\address{Max Planck Institute for the Physics of
  Complex Systems, Noethnitzer Str. 38, D-01187 Dresden, Germany}

\author{R. Everaers} \address{Universit\'e de Lyon, Laboratoire de
  Physique, \'Ecole Normale Sup\'erieure de Lyon, CNRS UMR 5672, 46
  all\'ee d'Italie, 69364 Lyon Cedex 07, France}
\address{Max Planck
  Institute for the Physics of Complex Systems, Noethnitzer Str. 38,
  D-01187 Dresden, Germany} 
\author{A.C. Maggs} 
\address{Laboratoire
  de Physico-Chimie Th\'eorique, UMR CNRS--ESPCI 7083, 10 rue
  Vauquelin, F-75231 Paris Cedex 05, France}

\end{center}
\end{titlepage}
\begin{abstract}
  In systems composed of water and hydrocarbons Van der
  Waals-interactions are dominated by the non-retarded, classical
  (Keesom) part of the Lifshitz-interaction; the interaction is
  screened by salt and extends over mesoscopic distances of the order
  of the size of the (micellar) constituents of complex fluids. We
  show that these interactions are included intrinsically in a
  recently introduced local Monte Carlo algorithm for simulating
  electrostatic interactions between charges in the presence of
  non-homogeneous dielectric media.
\end{abstract}
\section{Introduction}

In the special case of water--hydrocarbon systems, which notably
include biological systems, the weak optical contrast between water
and many hydrocarbons leads to Van der Waals interactions which are
dominated by the classical
(Keesom)-contribution~\cite{parsegian_ninham}.  Within the Lifshitz
formalism it is possible to perform analytical calculations only for
continuum descriptions of simple geometries such as planar interfaces
and lamellar structures~\cite{podgornik,ninham_parsegian,dean,netz}.
In the opposite extreme of atomistic Molecular Dynamics simulations,
the relevant partial charges on the water (solvent) molecules are
treated explicitely. This results in proper treatment of the Keesom
contribution of the Van der Waals interaction (because in the
microscopic simulation the microscopic dipoles are fluctuating).
However, current all-atom simulations are limited to the nanosecond
timescale, while the physical processes can take much longer.
Coarse-grained descriptions using implicit solvent models can help to
close this gap~\cite{markusdeserno}.  Typically, electrostatic
interactions are calculated from a (macroscopic) dielectric
theory~\cite{roux} while Van der Waals forces are modelled via
(effective) Lennard-Jones interaction with a cutoff of the order of
the ``grain size'' $\sigma$ of the coarse-grained
system~\cite{limbach}. However, neglecting the collective origin of
the Van der Waals interactions, namely electrostatic interactions
between fluctuating charge distributions, gives rise to several
problems:
\begin{itemize}
\item[--] in complex fluids Van-der-Waals interactions extend over
  distances which are comparable to the size of interacting
  objects~\cite{israelachvili,everaers_pre_03}. For polymers of
  amphiphilic systems organizing into micellar or lamellar structures,
  the characteristic length scale can be much larger than the size of
  constituting monomers.
\item[--] the results obtained by the approximated implicit solvent
  models are very sensitive to the value used for the dielectric
  constant, which turns out not to be a universal constant but simply
  a parameter that depends on the model used~\cite{warshel};
\item[--]the effect of screening by salt of a classical contribution
  to the Van-der-Waals interaction~\cite{mahanty_ninham} is not
  included.
\end{itemize}

Recent
publications~\cite{maggs_rossetto,maggs_auxfield,maggs_everaers}
reformulated the problem of electrostatic interactions between charges
in the presence of non-homogeneous dielectric media: Long ranged
electrostatic interactions between charges are generated {\it
  dynamically} via local interactions between charges, the medium and
the electric field.  Previously in~\cite{maggs_auxfield} the
interaction between small dielectric inhomogeneities was considered in
the dilute limit.  Here we generalize the proof to arbitrary
densitites and show that the method implicitly generates the many-body
effects in the zero frequency part of the Lifshitz interaction
regardless of the system density.

The paper is structured as follows: After a short review of Lifshitz
theory, in Section~\ref{sec:algorithm} we present the central result
of the paper - the theoretical basis of the simulation method and its
relation to Lifshitz theory. To be able to treat systems with general
geometries we have to validate our method for the case of simple
systems where the theoretical result can be used for the comparison.
Therefore we have chosen a triple slab system since for this
particular geometry one can develop and test a technique for correct
simulation and thermodynamic integration. In Sec.~\ref{sec:results} we
present our simulations and compare to analytic theory for the triple
slab geometry.

\section{Theoretical background}
\subsection{Lifshitz theory}
Dzyaloshinskii et al~\cite{lifshitz_dzyaloshinskii} recasted Van der
Waals forces in terms of interactions between continuous dielectric
media, mediated by the electromagnetic field. The result corresponds
to summing a series of multi-body interactions between fluctuating
charges.  In describing Van der Waals interactions the specificity of
the condensed medium is completely taken into account by using its
dielectric function $\epsilon(\omega)$, $\omega$ being the frequency
of electromagnetic field. In particular, the electromagnetic field
fluctuation free energy $\mathcal F$ is given by
\begin{equation}
  \mathcal F=k_BT\mathop{{\sum}'}_{n=0}^{\infty}\ln\mathcal D(i\xi_n)
\end{equation}
where $k_B$ is the Boltzmann constant and $T$ is the temperature. The
$n$ summation is over bosonic Matsubara frequencies $\xi_n=2\pi
nk_BT/\hbar$. The prime in the summation reflects the fact that the
$n=0$ term is taken with a weight $1/2$. The secular mode equation (or
dispersion equation) $\mathcal D(i \xi_n)=0$ gives the eigenfrequencies
of the electrodynamic field modes in the specified geometry.

For the case of two plane parallel half-spaces with dielectric constant
$\epsilon_1$ separated by the gap of length $l$ and dielectric constant
$\epsilon_2$ one can explicitely derive the free energy (per unit area) of the
interaction~\cite{mahanty_ninham}:
\begin{align}
  &\mathcal F(l)=\frac{k_BT}{8\pi l^2}\mathop{{\sum}'}_{n=0}^{\infty}I(\xi_n,l)\label{eq:general_lifshitz}\\
  \begin{split}
    I&(\xi_n,l)\equiv\\
    &\left(\frac{2\xi_nl\sqrt{\epsilon_2}}{c}\right)^2\int_1^{\infty}dp\, p\left(\ln\left[1-\Delta^2\exp\left(\frac{-2p\xi_nl\sqrt{\epsilon_2}}{c}\right)\right]\right.\\
    &+\left.\ln\left[1-\overline{\Delta}^2\exp\left(\frac{-2p\xi_nl\sqrt{\epsilon_2}}{c}\right)\right]\right)\\
  \end{split}\\
  \Delta&=\left(\frac{q\epsilon_2-p\epsilon_1}{q\epsilon_2+p\epsilon_1}\right),\;\overline{\Delta}=\left(\frac{q-p}{q+p}\right),
\; q=\sqrt{p^2-1+(\epsilon_1/\epsilon_2)}
\end{align}
The susceptibilities $\epsilon=\epsilon(i\xi_n)$ are evaluated on the
imaginary frequency axes.

The fluctuation driven electromagnetic interactions may be classical
or quantum in origin. Usually the temperatures of interest for
condensed media are low compared with $\hbar\omega_0$, where
$\omega_0$ is a typical frequency in the system which is often in the
ultraviolet ($T_0\sim\hbar\omega_0/k_B \sim 7\times 10^4
K$)~\cite{landau_VIII}.  In most condensed matter systems Van der
Waals interactions are thus dominated by quantum fluctuations.
Important exceptions occur in mixtures of polar liquids (e.g. water)
and hydrocarbon based (macro)molecules, a situation of considerable
interest for biological and biophysical problems. This is a
consequence of two effects. Firstly, there is low contrast between the
dielectric response of water and hydrocarbons in the optical part of
the spectrum.  Secondly, there is a {\em large} contrast at {\em
  low}-frequencies due to orientational fluctuations of dipoles in
polar liquids.

If one works in the gas phase, rather than in condensed media, and
considers the interaction energy between two water molecules in
vacuum, the corresponding classical {\it Keesom
  forces}~\cite{israelachvili} at room temperature are characterized
by the prefactor to the interaction in $1/r^6$:
$C_6^{Keesom}=96\times10^{-79}J\,m^6$ considerably larger than the
quantum (known as {\it dispersion}) contribution with the strength
$C_6^{disp}=33\times10^{-79}J\,m^6$ \cite{hirschfelder}. As a result
the zero-frequency contribution to the Van der Waals interaction in
water-hydrocarbon systems dominates and gives approximately $60\%$ of
the net interaction potential~\cite{israelachvili}.

When one drops, in the sum Eq.~\ref{eq:general_lifshitz}, the terms
for which $n\ne 0$ (which are essentially quantum mechanical) one
finds~\cite{mahanty_ninham}:
\begin{equation}\label{eq:zero_frequency_lifshitz}
  \mathcal F(l)_{n=0}= \frac{k_BT}{16\pi l^2}\int\limits_0^{\infty}x\,dx\,\ln\left\{1-\left(\frac{\epsilon_1(0)-\epsilon_2(0)}{\epsilon_1(0)+\epsilon_2(0)}\right)^2\mbox{e}^{-x}\right\}
\end{equation}

Eq.~\ref{eq:zero_frequency_lifshitz} can be derived using a different
approach~\cite{netz, dean}. Dean et al.~\cite{dean} have shown that if
one considers a thermal field theory for the field $\psi$ with purely
electrostatic Lagrangian
\begin{equation}
  \mathcal L[\psi]=\frac12\int\epsilon(\vect r)(\grad\psi)^2\,d^3\vect r
\end{equation}
the zero frequency Lifshitz term can be obtained from the partition
function of field $\psi$:
\begin{equation}
  Z=\int d[\psi]\exp(\beta\mathcal L[\psi])
\end{equation}
where $\beta=1/(k_BT)$. After changing in the latter formula the axes
of functional integration via $\psi\to -i\phi$ one recovers the
partition function of the dielectric system~\cite{dean}:
\begin{equation}\label{eq:functional_det}
  \begin{split}
    \mathcal Z&=\int d[\phi]\exp\left(\frac{\beta}2\int\phi\vect{\nabla}\epsilon(\vect r)\grad{\phi}\, d^3\vect r\right)\\
    &=[\mbox{det}(-\vect{\nabla}\epsilon(\vect r)\vect{\nabla})]^{-1/2}
  \end{split} 
\end{equation}
where $\mbox{det}(-\vect{\nabla}\epsilon(\vect r)\vect{\nabla})$ is
formally understood as the product of eigenvalues of operator
$-\vect{\nabla}\epsilon(\vect r)\vect{\nabla}~\cite{bailin}$. Finally
Eq.~\ref{eq:zero_frequency_lifshitz} can be recovered for the case of
planar geometry if one calculates the free energy from the usual
thermodynamic relation $\mathcal F=-k_BT\ln\mathcal Z$.

In all that follows we will consider only the zero frequency term (see
Eq.~\ref{eq:zero_frequency_lifshitz}) of Lifshitz interaction and will drop
the subscript $n=0$. Furthermore we will consider only static
susceptibilities and will not specify the frequency argument of
$\epsilon$.
\subsection{Triple slab geometry}
\begin{figure}[htbp]
  \centerline{
    \includegraphics[width=2.7in]{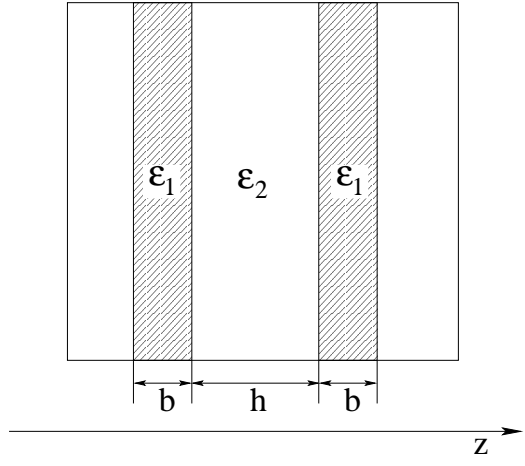}
  }
  \caption{Symmetric triple layer.} 
  \label{fig:triple_slab}
\end{figure}

A triple--slab geometry (Fig.~\ref{fig:triple_slab}) belongs also to
the class of analytically solvable geometries.  We will use it to
compare our simulations to known results.  It is also easy to treat in
periodic boundary conditions. As the free energy is an extensive
quantity~\cite{landau_lifshitz_stat_phys}, it contains a volume
contribution as well as a surface contribution. We are interested in
the distance dependence of the {\it surface} part of free energy. The
triple-slab geometry allows one to change the distance between two
slabs without any changes in the volume of dielectric materials in
finite systems. Hence the volume contribution in such a system can be
easily separated from the surface free energy we are interested in
(see Sec.~\ref{sec:results}).

One considers two slabs of material with dielectric constant
$\epsilon_1$ of thickness $b$ and area $L^2$ which are separated by a
distance $h$.  The dielectric constant of the external medium is given
by $\epsilon_2$ (see Fig.~\ref{fig:triple_slab}), so that
\begin{equation}\label{eq:eps_triple_slab}
  \begin{split}
    \epsilon(z)&=\epsilon_2+(\epsilon_1-\epsilon_2)\theta(z)+
(\epsilon_2-\epsilon_1)\theta(z-b)\\
    &+(\epsilon_1-\epsilon_2)\theta(z-b-h)+(\epsilon_2-\epsilon_1)\theta(z-2b-h)
  \end{split}
\end{equation}
where $\theta(z)$ is the Heaviside step function. 

The interaction energy per unit area is given by
\cite{ninham_parsegian,dean}:
\begin{equation}\label{eq:tripleslab_exactenergy}
  \begin{split}
  \mathcal F(h,b)&=\\
  &\frac{k_BT}{4\pi}\int_0^{\infty}dp\,p\,\mbox{ln}\left[1-\frac{\Delta^2(1-e^{-2bp})^2e^{-2hp}}{(1-\Delta^2e^{-2bp})^2}\right]
  \end{split}
\end{equation}
where
\begin{equation}
  \Delta = \frac{\epsilon_2-\epsilon_1}{\epsilon_2+\epsilon_1}
\end{equation}

One can also consider the pairwise approximation to the general result given
by Eq.~\ref{eq:tripleslab_exactenergy}. The major contribution to the integral
with respect to $p$ comes from the saddle point $p\approx 0$. Hence we can
write, since $\Delta^2<1$,
\begin{equation}
  \mbox{ln}(1-\Delta^2e^{-2hp})\simeq -\Delta^2e^{-2hp},
\end{equation}
and carry out the integration with respect to $p$ to find~\cite{ninham_parsegian}
\begin{equation}\label{eq:pairwise_approx}
  \mathcal F(h,b)\simeq -\frac{k_BT}{16\pi}\left(\frac{\Delta^2}{h^2}+
\frac{\Delta^2}{(h+2b)^2}+\frac{2\Delta^2}{(h+b)^2}\right)
\end{equation}
In the case of two infinite slabs $b\to\infty$ we have the usual result of
Hamaker theory~\cite{hamaker}:
\begin{equation}
  \mathcal F_{H}(h,b\to\infty)=-\frac{k_BT\Delta^2}{16\pi h^2}
\equiv -\frac{A}{12\pi h^2}
\end{equation}
where $A$ is the classical part of the Hamaker constant.
\section{A local Monte Carlo algorithm for generating Van der Waals interactions}\label{sec:algorithm}
In the following we describe our simulation method. At zero
temperature the Coulomb interaction results from minimizing the energy
\begin{equation}\label{eq:electrostat_energy}
  \mathcal U=\frac12\int\frac{\vect D^2}{\epsilon(\vect r)}d^3\vect r
\end{equation}
where $\epsilon(\vect r)$ is assumed isotropic and $\vect D$ is the
electric displacement constrained by Gauss´ law, $\diverg D = \rho$;
$\rho$ is the external charge density.  We assume throughout the paper
that the dielectric constant of the vacuum is $\epsilon_0=1$. This
constrained minimization problem for $\vect D$ can be solved with the
help of a Lagrange multiplier $\phi(\vect r)$ by looking for
stationary points of the functional $\vect
D$~\cite{Schwinger/deRaad:1998},
\begin{equation}\label{eq:U0_functional}
  \mathcal U[\vect D]=\int\left\{\frac{\vect D^2}{2\epsilon(\vect r)}-\phi(\vect r)(\diverg D(\vect r)-\rho(\vect r))\right\}d^3\vect r
\end{equation}
and is given by
\begin{equation}\label{eq:energy_phi}
  \mathcal U_0=\frac 12\int\epsilon(\vect r)(\grad\phi)^2\,d^3\vect r
\end{equation}
$\phi$ is the solution of the Poisson equation
\begin{equation}\label{eq:poisson}
  \diverg(\epsilon(\vect r)\grad \phi)=-\rho
\end{equation}

The true interest of the formulation appears in Monte Carlo since one
can generate local dynamic systems which sample the partition function
\begin{equation}\label{eq:partition_function}
  \mathcal Z=\int\prod_{i=1}^{N}d^3\vect r_i\prod_{\vect r}
\mathcal D\vect D(\vect r)\delta(\diverg D-\rho(\vect r))
e^{-\frac{\beta}{2}\int\vect D^2/\epsilon(\vect r)d^3\vect r}
\end{equation}
Following~\cite{maggs_rossetto} we discretize the system placing
particles on a simple cubic lattice with vector fields such as ${\vect
  D} $ on the links.  This formulation is numerically efficient
because a local variation in $\rho$ requires only a local update of
the field ${\vect D}$. For problems involving dielectric
inhomogeneities (macroparticles with dielectric constant differing
from the surroundings) one has to choose an appropriately interpolated
value of the dielectric function.  The dielectric function is placed
also on the link (Ref.~\cite{maggs_auxfield}) and is given by the
harmonic average
\begin{equation}\label{eq:interpolationf_dielectric}
  \frac{2}{\epsilon_{n\mu}}=\frac{1}{\epsilon_n}+\frac{1}{\epsilon_{n+\mu}}
\end{equation}
where $\{n\mu\}$ is the link which goes from the site $n$ in the
$\mu$--direction, $\mu=1,\,2,\,3$.
In the absence of charged particles the partition function
Eq.~\ref{eq:partition_function} becomes
\begin{equation}
  \mathcal Z=\int\prod_{\vect r}\mathcal D
\vect D(\vect r)\delta(\diverg D)e^{-\frac{\beta}{2}\int\vect D^2/\epsilon(\vect r)d^3\vect r}
\end{equation}
Introducing an auxiliary field $\phi$ to implement the Gauss' law constraint
and using the identity $2\pi\delta(x)=\int e^{i\phi x}\, d \phi$ the last
equation is equivalent to
\begin{equation}\label{eq:maggs_partition_func}
  \begin{split}
    \mathcal Z &= C_1\int\prod_{\vect r}d\phi\prod_{\vect r}\mathcal 
D\vect D e^{i\int\phi\diverg Dd^3\vect r-
\frac{\beta}{2}\int\vect D^2/\epsilon(\vect r)d^3\vect r}\\
    &= C_1\int\prod_{\vect r}d\phi\prod_{\vect r}
\mathcal D\vect D e^{-i\int\vect D\grad\phi d^3\vect r-
\frac{\beta}{2}\int\vect D^2/\epsilon(\vect r)d^3\vect r}\\
    &=C_2\prod_{\vect r}\epsilon(\vect r)^{3/2}
\int\prod_{\vect r}d\phi e^{-\frac{1}{2\beta}
\int\epsilon(\vect r)\,(\grad\phi)^2d^3\vect r}\\
    &=C_3\left[\mbox{det}(-\vect{\nabla}
\cdot\epsilon(\vect r)\vect{\nabla})\right]^{-1/2}
  \end{split}
\end{equation}
where the constants $C_1$, $C_2$ and $C_3$ are of no further
interest~\footnote{In the case of moving particles the variations of
  the term $\sqrt{\epsilon}$ can add nontrivial contributions to the
  contact energy.}.  Comparing Eq.~\ref{eq:maggs_partition_func} and
Eq.~\ref{eq:functional_det} we conclude that our method produces the
zero-frequency term of the Lifshitz interaction. One
  has to note, in spite of the fact that intermediate expressions in
  deriving Eq.~\ref{eq:maggs_partition_func} contain complex
  contributions, the algorithm directly samples the real constrained
  partition function given by Eq.~\ref{eq:partition_function}.
\section{Numerical validation}\label{sec:results}
We simulate a triple slab system Fig.~\ref{fig:triple_slab} with two
values of dielectric constant of the media: $\epsilon_1=2$ and
$\epsilon_1=78$ using different values of the lattice constant $a$. The
dielectric constant of the intermediate region is $\epsilon_2=1$ for
both cases. The size of the box is $L=15.0$, the width of the slab
$b=1.0$. 

In order to calculate the free energy we perform a thermodynamic integration
\cite{frenkel_smit}.  For the reference system (denoted by $I$) the
uniform dielectric constant $\epsilon_2=1.0$ is assigned.We sample the system with
the potential energy $\mathcal U$ which depends linearly on the coupling
parameter $\lambda$:
\begin{equation}\label{eq:energy_lambda}
  \begin{split}
  \mathcal U(\lambda)&=(1-\lambda)\mathcal U_{I}+\lambda\mathcal U_{II}\\
  &=(1-\lambda)\int d^3\vect r\frac{\vect D^2}{2\epsilon_2}+
  \lambda\int d^3\vect r\frac{\vect D^2}{2\epsilon(\vect r)}
  \end{split}
\end{equation}
For $\lambda=1$ we recover our system of interest (denoted by $II$).  The
system with energy $U(\lambda)$ is equivalent to the system with the following
dielectric function:
\begin{equation}
  \epsilon_t(\lambda,\vect r)=\frac{\epsilon(\vect r)\epsilon_2}{\epsilon(\vect r)+\lambda(\epsilon_2-\epsilon(\vect r))}
\end{equation}
Finally, the free energy difference between systems $II$ and $I$ can
be found from the following expression:
\begin{equation}\label{eq:therm_freeenergy}
  \mathcal F(II)-\mathcal F(I)=\int_0^1d\lambda\left<\frac{\partial\mathcal U}{\partial\lambda}\right>_{\lambda}
\end{equation}
where $\left<...\right>_{\lambda}$ denotes an ensemble average for a
system with energy Eq.~\ref{eq:energy_lambda}.

The numerical calculation of a free energy is always demanding.  We
have approximated the integral in Eq.~\ref{eq:therm_freeenergy} by a
summation over 20 intervals in $\lambda$. The fluctuating field
$\vect D$ is sampled by a worm algorithm~\cite{levrel_pramana, alet}.
Simulation at each $\lambda$ point involved an equilibration period of
$800$ sweeps, where a sweep is 20 worms, followed by a consequent run
of another $800$ sweeps configurations. The error bars and average
values of free energy have been calculated from $500$ values of free
energy. Simulations were performed on an AMD Opteron 2.4GHz processor.
Total simulation time for a one measured point was 2 days for $a=1.0$
and 24 days for $a=0.5$.

The free energy calculated in this way gives the full contribution
which includes self--energies of individual slabs as well as the
interaction energy between slabs. In contrast
Eqs.~\ref{eq:pairwise_approx} and~\ref{eq:tripleslab_exactenergy}
represent only the interaction part of the excess free energy. In a
system with periodic boundary conditions it is difficult to calculate
the limit $h\to\infty$ which corresponds to calculating the
self--energy part. Therefore we perform an interpolation of our
simulation results by the function Eq.~\ref{eq:pairwise_approx} and
extrapolate them to the region $h\to\infty$ to find the asymptotic
value. Further we subtract this contribution from the total free
energy Eq.~\ref{eq:therm_freeenergy}. Of course such a procedure leads
to small deviations from the analytic curve which can be clearly seen
on the corresponding plots.
%
\begin{figure}[tbp]
  \centerline{
    \includegraphics[width=4.in,clip]{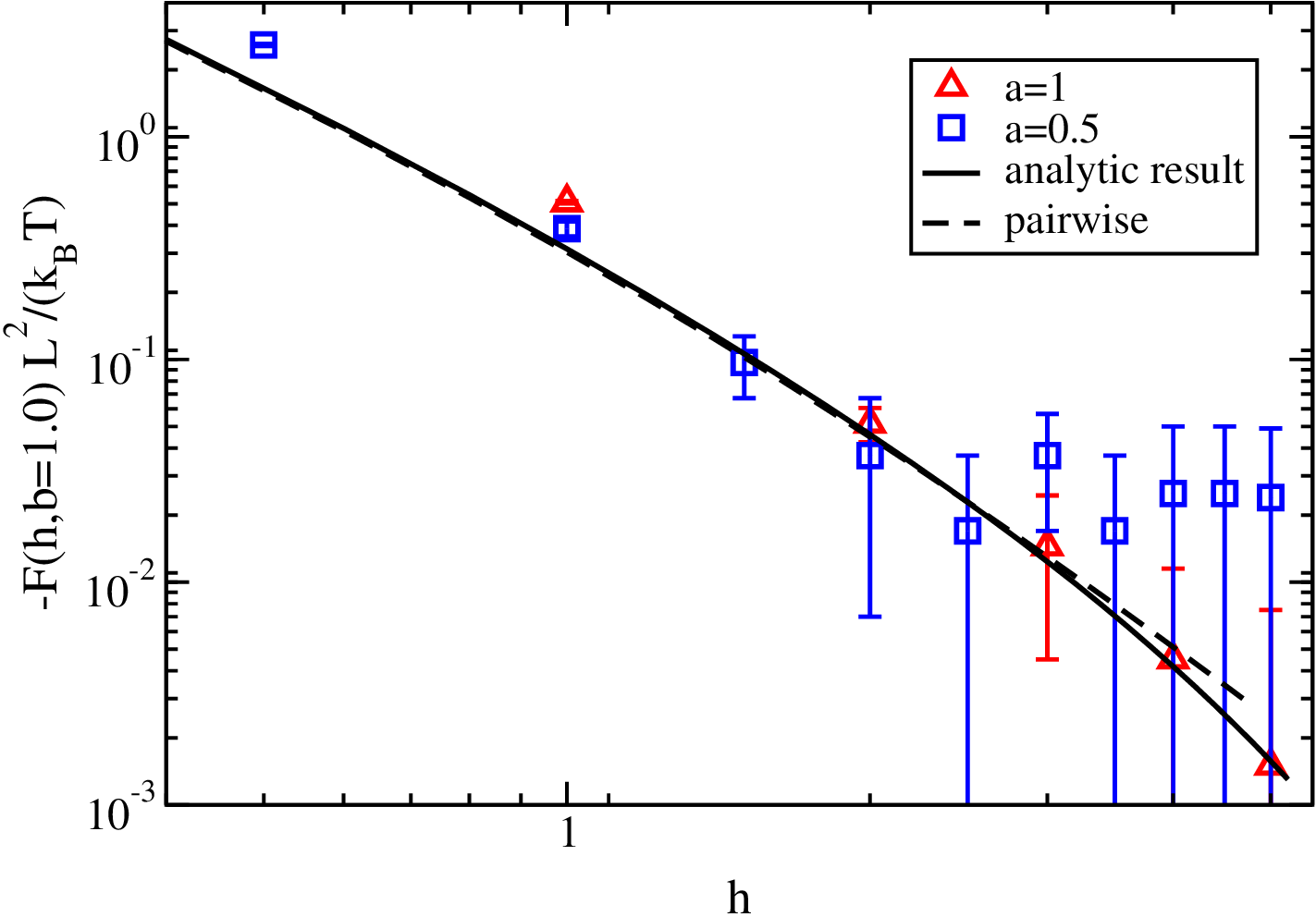}
  }
  \caption{ \label{fig:epsilon_2}Free energy of the slab at $\epsilon_1=2$.}
  \vspace*{0.55cm}
  \centerline{
    \includegraphics[width=4.in,clip]{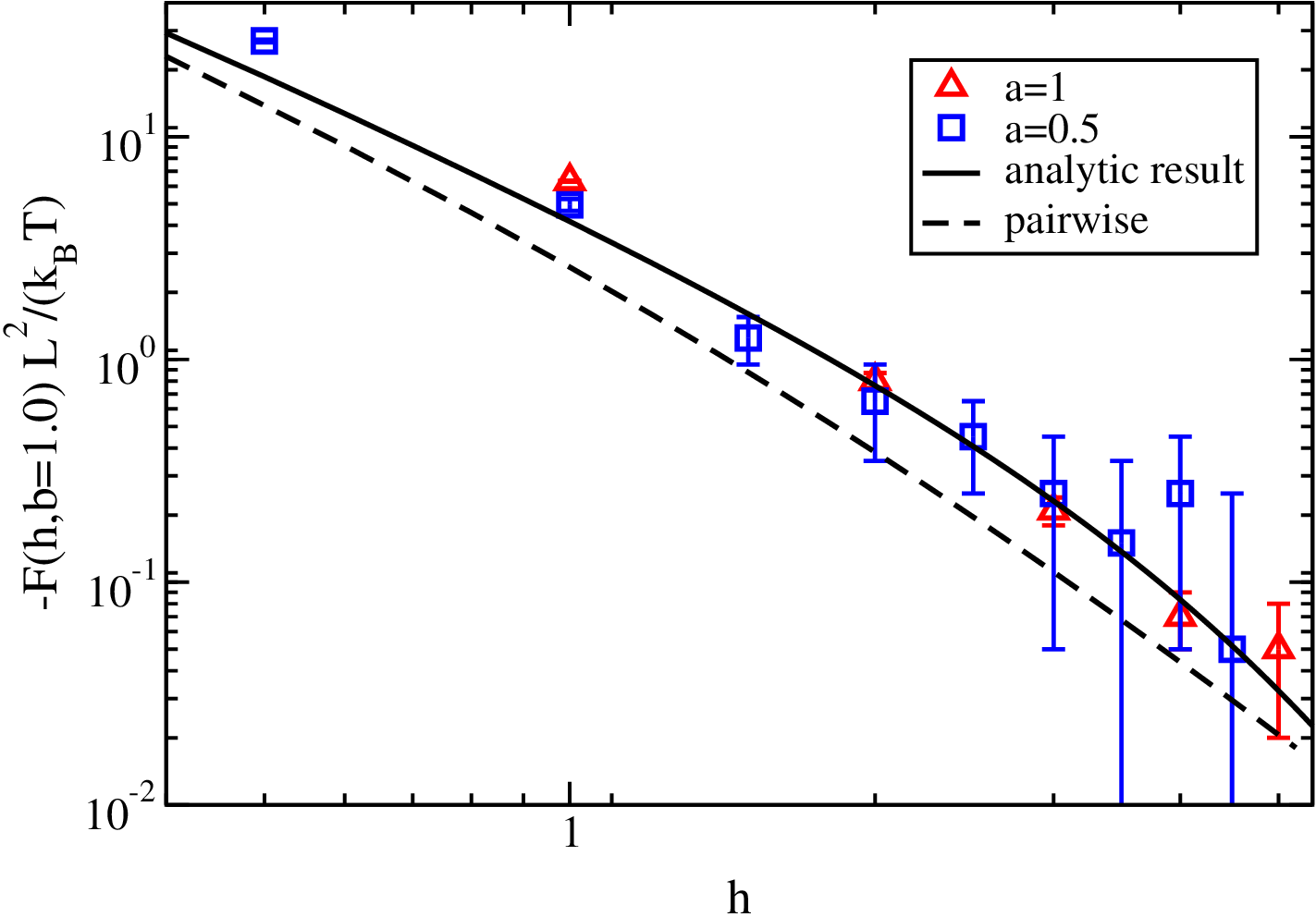}
  }
  \caption{\label{fig:epsilon_78}Free energy of the slab at $\epsilon_1=78$.}   
\end{figure}
We are interested in observing the free energy of the system as we
change the separation between slabs. Our results are shown in 
Fig.~\ref{fig:epsilon_2} for
$\epsilon_1=2.0$ and in Fig.~\ref{fig:epsilon_78} for
$\epsilon_1=78.0$. 
In both cases, a comparison of results for  two
different values of the lattice constant $a=0.5;\,1.0$ shows 
that the errors due to lattice discretization are small.
In particular, the data reproduce the analytic result
Eq.~\ref{eq:tripleslab_exactenergy} which differs significantly from 
the pairwise (Hamaker) curve Eq.~\ref{eq:pairwise_approx} for large
dielectric contrasts.
\section{Conclusion}
We have shown that a recent Monte Carlo algorithm for the simulation
of electrostatic interactions in heterogeneous dielectric media
implicitly generates the zero-frequency part of the Lifshitz
interaction including all many-body effects. The interactions make the
dominant contribution to the Van der Waals attraction in
hydrocarbon-water systems as they are typically found in soft and
biological condensed matter~\cite{israelachvili}. The method is easily
applicable to systems with interfaces and spatially varying dielectric
constants of arbitrary geometry and allows the inclusion of fixed and
free charges.
\section*{ACKNOWLEDGMENT}
We would like to thank Lucas Levrel and Samuela Pasquali for
discussions.  The work was supported by the Volkswagenstiftung. RE is
supported by a chair of excellence grant from the Agence Nationale de
Recherche (France).
\bibliographystyle{unsrt} 
\bibliography{papers}

\begin{thebibliography}{10}

\bibitem{parsegian_ninham}
V.A. Parsegian and B.W Ninham.
\newblock {Temperature-Dependent van der Waals Forces}.
\newblock {\em Biophysical Journal}, 10:664, 1970.

\bibitem{podgornik}
Podgronik R., R.H. French, and Parsegian V.A.
\newblock {Nonadditivity in van der Waals interactions within multilayers}.
\newblock {\em Journal of Chemical Physics}, 124:044709, 2006.

\bibitem{ninham_parsegian}
B.W Ninham and V.A. Parsegian.
\newblock {van der Waals Forces across Triple-Layer Films}.
\newblock {\em Journal of Chemical Physics}, 52(9):4578, 1970.

\bibitem{dean}
D.S. Dean and R.R. Horgan.
\newblock Electrostatic fluctuations in soap films.
\newblock {\em Physical Review E}, 65:061603, 2002.

\bibitem{netz}
R.R. Netz.
\newblock {Static van der Waals interaction in electrolytes}.
\newblock {\em European Physical Journal E}, 5(189), 2001.

\bibitem{markusdeserno}
B.J. Reynwar, G.~Illya, V.A. Harmandaris, M.M. Muller, K.~Kremer, and
  M.~Deserno.
\newblock Aggregation and vesiculation of membrane proteins by
  curvature-mediated interactions.
\newblock {\em Nature}, 447:461--464, 2007.

\bibitem{roux}
B.~Roux and T.~Simonson.
\newblock Implicit solvent models.
\newblock {\em Biophyiscal Chemistry}, 78:1--20, 1999.

\bibitem{limbach}
H.J. Limbach and C.~Holm.
\newblock {Single-Chain Properties of Polyelectrolytes in Poor Solvent}.
\newblock {\em J. Phys. Chem. B.}, 107:8041--8055, 2003.

\bibitem{israelachvili}
J.N. Israelachvili.
\newblock {\em {Intermolecular and Surface Forces}}.
\newblock Academic Press, 1992.

\bibitem{everaers_pre_03}
R.~Everaers and M.R. Ejtehadi.
\newblock Interaction potentials for soft and hard ellipsoids.
\newblock {\em Physical Review E}, 67:041710, 2003.

\bibitem{warshel}
C.~Schutz and A.~Warshel.
\newblock What are the dielectric "constants" of proteins and how to validate
  electrostatic models?
\newblock {\em PROTEINS-STRUCTURE FUNCTION AND GENETICS}, 44:400, 2001.

\bibitem{mahanty_ninham}
J.~Mahanty and B.W. Ninham.
\newblock {\em {Dispersion Forces}}.
\newblock Academic Press, London, 1976.

\bibitem{maggs_rossetto}
A.C. Maggs and V.~Rossetto.
\newblock {Local Simulation Algorithms for Coulomb Interactions}.
\newblock {\em Phys. Rev. Lett.}, 88:196402, 2002.

\bibitem{maggs_auxfield}
A.C. Maggs.
\newblock {Auxiliary field Monte Carlo for charged particles}.
\newblock {\em Journal of Chemical Physics}, 120:3108--3118, 2004.

\bibitem{maggs_everaers}
A.C. Maggs and R.~Everaers.
\newblock Simulating nanoscale dielectric response.
\newblock {\em Physical Review Letters}, 96:230603, 2006.

\bibitem{lifshitz_dzyaloshinskii}
I.E. Dzyaloshinskii, E.M. Lifshitz, and L.P. Pitaevski.
\newblock {The general theory of van der Waals forces}.
\newblock {\em Advances in Physics}, 10(38):165, 1961.

\bibitem{landau_VIII}
L.D. Landau and E.M. Lifshitz.
\newblock {\em Electrodynamics of continuous media}.
\newblock Pergamon, 1998.

\bibitem{hirschfelder}
J.O. Hirschfelder, C.F. Curtiss, and R.B. Bird.
\newblock {\em Molecular theory of gases and liquids}, chapter~13, page 988.
\newblock New York, Wiley, 1964.

\bibitem{bailin}
D.~Bailin and A.~Love.
\newblock {\em Introduction to gauge field theory}.
\newblock IOP Publ., 1993.

\bibitem{landau_lifshitz_stat_phys}
L.D. Landau and E.M. Lifshitz.
\newblock {\em {Statistical Physics (Course of Theoretical Physics, Volume
  5)}}.
\newblock Butterworth-Heinemann, 2000.

\bibitem{hamaker}
H.C. Hamaker.
\newblock {The London - Van Der Waals attraction between spherical particles}.
\newblock {\em Physica}, 4:1058, 1937.

\bibitem{Schwinger/deRaad:1998}
J.~Schwinger, L.L. DeRaad, K.A. Milton, and Wu-yang Tsai.
\newblock {\em {Classical Electrodynamics}}.
\newblock Perseus Books, 1998.

\bibitem{frenkel_smit}
D.~Frenkel and B.~Smit.
\newblock {\em {Understanding Molecular Simulation}}.
\newblock {Academic Press}, 2002.

\bibitem{levrel_pramana}
L.~Levrel, F.~Alet, J.~Rottler, and A.C. Maggs.
\newblock {Local Simulation Algorithms for Coulombic Interactions}.
\newblock {\em PRAMANA -- journal of physics}, 64:1001, 2005.

\bibitem{alet}
F.~Alet and E.S. Sorensen.
\newblock {Cluster Monte Carlo algorithm for the quantum rotor model}.
\newblock {\em Physical Review E}, 67:015701, 2003.

\end{thebibliography}
\end{document}